\documentclass[twocolumn]{article}
\usepackage{icqproc,epsf,amssymb,amsmath,epsfig}
\begin{document}
\thispagestyle{empty}

\twocolumn[
%\vspace*{30mm}
\begin{footnotesize}
\begin{flushright}

7th International Conference on Quasicrystals (ICQ7), Stuttgart
1999,

to be published in Materials Science and Engineering A
\end{flushright}
\end{footnotesize}
\vspace*{15mm}
\begin{LARGE}
\begin{center}
%
%  TITLE
%
Conditions on the Occurrence of Magnetic Moments in Quasicrystals
and Related Phases
%
%  END TITLE
%
\end{center}
\end{LARGE}
\begin{large}
\begin{center}
%
%  AUTHORS
%
Guy Trambly de Laissardi\`ere$^{a}$, Didier Mayou$^{b}$
%
%  END AUTHORS
%
\end{center}
\end{large}
\begin{footnotesize}
\begin{it}
\begin{center}
%
%  ADDRESS
%
a) Laboratoire de Physique Th\'eorique et Mod\'elisation,
Universit\'e de Cergy-Pontoise, Neuville, 95031 Cergy-Pontoise,
France; and Laboratoire L\'eon Brillouin, CEA-CNRS Saclay,
France.\\ b) Laboratoire d'Etudes des Propri\'et\'es Electronique
des Solides (CNRS), 38042 Grenoble C\'edex 9, France.
%
%  END ADDRESS
%
\end{center}
\end{it}
\end{footnotesize}
\begin{footnotesize}
\begin{center}
%
%  DATE
%
1999/10/12 %received in revised form
%
%  END DATE
%
\end{center}
\end{footnotesize}
\vspace{4ex}
\begin{small}
\hrule\vspace{3ex}
\begin{minipage}{\textwidth}
{\bf Abstract}\vspace{2ex}\\
\hp
%
%  ABSTRACT
%
We analyze the criterion for the appearance of magnetic moments in
quasicrystals approximants and liquids Al-Mn. In a Hartree-Fock
scheme, it is shown that the Stoner criterion ($Un_d(E_F) \geq 1$)
does not apply to these systems due to the presence of the
pseudogap in the density of states. We give an generalized
criterion that can take into account the particular density of
state. The main result is that the appearance of a magnetic moment
on a Mn atom depends strongly on its position in the crystalline
structure. Indeed some particular positions of the Mn atoms lead
to magnetic moments whereas other positions are more stable when
the Mn is non-magnetic. This criterion is used to analyzed the
electronic structure of two approximants (1/1-Al-Pd-Mn-Si and
T-Al-Pd-Mn) calculated by the self-consistent LMTO-ASA method.
%
%  END ABSTRACT
%
\vspace{2.5ex}
\end{minipage}\vspace{3ex}
\hrule
\end{small}\vspace{6ex}
]

%
%  MAIN TEXT
%

%%%%%%%%%%%%%%%%%%%%%%%%%%%%%%%%%%%%%%%%%%%%%%%%%
%%% Section 1
%%%
\section{Introduction}

\hp

The magnetic properties of quasicrystals have been studied since
the discovery of icosahedral Al-Mn phase. The main focus of this
article is the existence of local moments carried by Mn atoms. In
crystals ($\rm Al_6Mn$, $\rm Al_{12}Mn$...) and some approximants
($\alpha$- and $\beta$-Al-Mn-Si...) the Mn atoms do not carry a
magnetic moment (``non-magnetic'' atoms, spin $\rm gS=m=0$). In
icosahedral Al-Mn, Al-Pd-Mn and many of their approximants
($\mu$-$\rm Al_4Mn$, 1/1-Al-Pd-Mn-Si, T-Al-Pd-Mn...), most of the
Mn atoms are non-magnetic but few Mn atoms carry a magnetic moment
(``magnetic'' atoms, $\rm m\neq0$). By contrast, in the liquid in
equilibrium with these phases, there is a large proportion of
magnetic Mn atoms. See
Refs.\;\cite{Francoise_Approx,Francoise_Liquide} and Refs. in
them.

In Al-Mn alloys, Mn atoms are closed to the magnetic /
non-magnetic transition. The magnetic state of Mn depends on the
local density of states (local DOS) on Mn
atoms\;\cite{GuyEuro93,Hafner_Mag}. These local DOS is influenced
by the local environment and we show that it depends also on the
environment up to $\rm 5-10\;\AA$. The usual criterion to study
the magnetic states of transition metal (TM) atom in non-magnetic
host is the Friedel-Anderson criterion (equivalent to the Stoner
criterion for a TM impurity in a non-metallic host). This
criterion assumes that sp valence states are free states, but in
crystals and quasicrystals that is no more valid and it is
necessary to generalize the criterion.

%%%%%%%%%%%%%%%%%%%%%%%%%%%%%%%%%%%%%%%%%%%%%%%%%
%%% Section 2
%%%
\section{Generalized Stoner criterion}

The magnetic state of a TM atom in aluminum host results from the
competition between the Coulomb interaction and the kinetic energy
of the d electrons. The Coulomb interaction is $\rm {\cal
E}_{C}=UN_{d\downarrow}N_{d\uparrow}= \frac{U}{4}(N_d^2-m^2)$,
where N$_d$ (N$_d$ $=$ N$_{d\uparrow}+$ N$_{d\downarrow}$) is the
total number of d electrons and m (m $=$ N$_{d\uparrow}-$
N$_{d\downarrow}$) is the local magnetic moment on the TM atom.
${\cal E}_{C}$ decreases when m increases, so it favors the
occurrence of a magnetic moment (m $\neq 0$). The kinetic energy
(or band energy), $\Delta$, of the d electrons increases when the
hybridization between the d orbital and the sp nearly free states
increases. Roughly speaking $\Delta$ is proportional to width of
the local d DOS. Therefore, the understanding of magnetic moment
requires to analyze carefully the local DOS. In the Al-Mn alloys,
because of the strong sp-d hybridization
\cite{FujiPRL91,BelinEuro94,GuyPRB95,Hafner97}, the local DOS
depends widely on the atomic structure: the local environment
(number and chemical nature of the neighbors, local symmetry) must
be important, but it depends also on the medium range order
because the sp states are delocalized. This last effect is the
most important for Mn atoms as respect with other TM atoms (Cr,
Fe, Co, Ni), because the d band of Mn is half filled. In the
following we analyze this effect.

Let us recall first the case of one Mn impurity in a non-metallic
Al host described by the well known Virtual Bound States
model\;\cite{Friedel_Anderson} for which one d orbital is coupled
with the sp free states. The result of this model is independent
of the position of the Mn atom in the alloy. The Mn d DOS is a
Lorentzian that satisfies the rigid band model, i.e. the d DOS is
shifted when the on site energy, E$_d$, varies but its shape does
not change. In the Hartree approximation, the energy E$_d$ is
obtained by the self consistent relation:
\begin{eqnarray}
E_{d,\sigma} = E_d^0 + U(N_{d,-\sigma} - N_{d,-\sigma}^0)\;.
\label{EqHF}
\end{eqnarray}
N$_{d\sigma}$, and N$_{d\sigma}^0$ are the number of d electrons
for a Mn atom with spin $\sigma$ in the alloys, and for an
isolated Mn atom, respectively. From the rigid band model, the
criterion of the appearance of a localized moment is easy to
calculate by writing these self consistent equations for the spin
up and the spin down. It follows that a localized moment exists if
(local Stoner criterion) \;\cite{Friedel_Anderson}:
\begin{eqnarray}
Un_d(E_F) \geq 1\;. \label{EqStoner}
\end{eqnarray}
For Mn atoms in a metallic host we have to consider 5 degenerated
d orbitals per Mn impurity. The exchange integral J between two
electrons, localized on two different orbitals of the same Mn
impurity, is taken into account by changing U to (U$-4$J).

As shown experimentally and theoretically the DOS of the
quasicrystals and their approximants is characterized by the
presence of a well defined pseudogap in the vicinity of the Fermi
energy (E$_F$) which contributes to their stabilization
(Hume-Rothery mechanism)\;\cite{Friedel88,FujiPRL91,GuyPRB95}. The
band energy is minimized when the number of valence electrons is
such that the Fermi sphere touches a ''predominant Brillouin
zone'', constructed by the Bragg planes located in the vicinity of
the Fermi sphere. The diffraction by Bragg planes splits the
valence sp states into bonding and antibonding states. That leads
to the formation of a valley, called pseudogap, in the partial sp
DOS near E$_F$. The energies of bonding (antibonding) states are
mainly below (above) the pseudogap. Because of the diffraction by
Bragg planes, there is no more translation invariance of valence
state amplitudes. Therefore the hybridization between the nearly
free sp states and a d orbital depends strongly on the Mn
position, {\bf r}$_d$, of the Mn atom in the alloys. In
particular, it has been shown that the partial d DOS on Mn
exhibits a pseudogap or not, depending on the Mn
position\;\cite{GuyEuro93}. Schematically, one find that there are
two extreme cases depending on {\bf r}$_d$ via the parameter
$\alpha({\bf r}_d)$, $\alpha({\bf r}_d)=\langle \cos({\bf K}.{\bf
r}_d+\Phi) \rangle$. Here, {\bf K} are the wave vectors of the
reciprocal lattice that define the Bragg plane from which the
pseudo-Brillouin zone is constructed, and $\Phi$ is a phase
depending on the lattice. If $\langle \cos ({\bf K}.{\bf r}_d +
\Phi) \rangle =1$, the amplitude of the bonding valence states is
minimized at {\bf r}~$=$~{\bf r}$_d$, and the amplitude of the
antibonding valence states is maximized at {\bf r}~$=$~{\bf
r}$_d$. Thus the d orbital is more coupled with the antibonding
valence states. On the other hand, if $\langle \cos ({\bf K}.{\bf
r}_d +\Phi)\rangle =-1$, the d orbital is more coupled with the
bonding valence states.

Moreover due to the large diffraction by Bragg plane, the rigid
band model hypothesis for the local d DOS is no more valid. The
Friedel-Anderson criterion for the appearance of a localized
moment (equation\;(\ref{EqStoner})) has to be generalized. The
revised criterion, calculated from the equations\;(\ref{EqHF}),
is:
\begin{eqnarray}
-U~\frac{\partial N_d}{\partial E_d} \geq 1\;, \label{EqCritere}
\end{eqnarray}
where N$_d$ is the number of d electrons of Mn in the non-magnetic
case.

\begin{figure}[t]
%\vspace{5 cm} %
\hfil \psfig{file=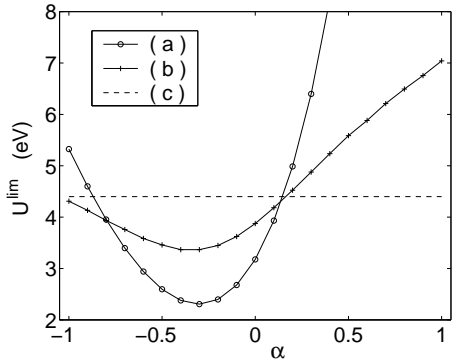,width=6cm} %
\caption{Criterion for the appearance of a localized magnetic
moment on Mn in Al host (see text). $\alpha= \langle \cos ({\bf
K}.{\bf r}_d + \Phi)\rangle$. With diffraction by Bragg plane: (a)
Stoner criterion, U$^{lim}=1/$n$_d$(E$_F$). (b) generalized
criterion, U$^{lim}=-1/(\partial$N$_d/\partial$E$_d)$. (c) without
diffraction by Bragg planes (Virtual Bound State hypothesis),
U$^{lim}=-1/(\partial$N$_d/\partial$E$_d)=1/$n$_d$(E$_F$).}
\label{FigCritere}
\end{figure}

One defines the energy U$^{lim}$ as the interaction energy above
which the Mn atom is magnetic (Fig.\,\ref{FigCritere}). The first
difference with the Virtual Bound State is that U$^{lim}$ varies
with the position {\bf r}$_d$ via the parameter $\langle \cos
({\bf K}.{\bf r}_d + \Phi) \rangle$.

%%%%%%%%%%%%%%%%%%%%%%%%%%%%%%%%%%%%%%%%%%%%%%%%%
%%%%%% Figure 1
%%%%%%
%%%%%%%%%%%%%%%%%%%%%%%%%%%%%%%%%%%%%%%%%%%%%%%%%

The calculated U$^{lim}$ gives also a satisfying model to
understand the difference between the magnetism of Al-Mn crystals,
quasicrystals and liquids. Whereas most of Mn atoms are
non-magnetic in crystals and quasicrystals, it has been found from
ab initio calculations\,\cite{Bratkovsky95} and experimental
measurements\,\cite{Francoise_Liquide} that the liquid Al-Mn
exhibit a large amount of localized magnetic moments. In the
crystals, d orbitals of stable Mn atoms are preferentially coupled
with antibonding sp states, $\langle \cos ({\bf K}.{\bf r}_d +
\Phi) \rangle \simeq 1$ (see Refs.\;\cite{GuyEuro93,GuyPRB95} and
next section); U$^{lim}$ is given by the curve
Fig.\,\ref{FigCritere}b which takes into account the diffraction
by Bragg planes. But in liquid, the topological disorder destroys
the diffraction by Bragg plane and U$^{lim}$ is given by the curve
Fig.\;\ref{FigCritere}c. As the temperature increases through the
fusion, the U$^{lim}$ decreases, and the Mn should go from a
non-magnetic state to a magnetic state.

%%%%%%%%%%%%%%%%%%%%%%%%%%%%%%%%%%%%%%%%%%%%%%%%%
%%% Section 3
%%%
\section{Energetic criterion}
\label{SecEnergie}

\begin{figure}[t]
%\vspace{5 cm}%
\hfil \psfig{file=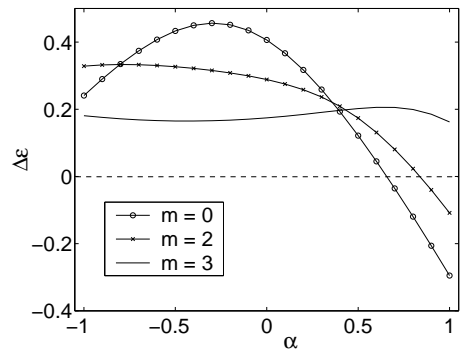,width=6cm} %
 \caption{ Variation $\Delta{\cal E}$ (eV) of the energy
due to one Mn d impurity in a metallic host as function of
$\alpha= \langle \cos ({\bf K}.{\bf r}_d + \Phi)\rangle$. Full
line, with diffraction by Bragg planes. Dashed line, without
diffraction by Bragg planes (impurity case).}
\label{FigModelEnergie}
\end{figure}

The generalized criterion presented in the last section can be
analyzed in terms of band energy. We calculated the energy of Mn
atoms in low concentration in Al host
(Fig.\,\ref{FigModelEnergie}), assuming a important diffraction by
Bragg plane. This model is described elsewhere\,\cite{GuyEuro93}.

Our analysis shows that the Mn state can be understood by simple
arguments as follows. If the Mn atom is non-magnetic its energy
varies with its position, {\bf r$_d$}, but the energy of the Mn
atoms with a high moment does not vary with {\bf r$_d$}. Indeed
the magnetic Mn atoms are less influenced by their environment
because their d bands are not close to E$_F$ when in fact the most
delocalized electron are electrons with energy E$_F$. In the
studied alloys the hybridization between the atoms is important
and the electronic energy of the Mn atoms is minimized for the Mn
positions {\bf r$_d$} satisfying $\langle \cos ({\bf K}.{\bf r}_d
+\Phi)\rangle \simeq 1$. In that case, the Mn is non magnetic and
the d DOS calculation exhibits\,\cite{GuyEuro93} a pseudogap near
E$_F$. By contrast the Mn atomic positions ($\langle \cos ({\bf
K}.{\bf r}_d +\Phi)\rangle \simeq -0.3$), that are less stable,
lead to magnetic Mn and no pseudogap near E$_F$ in the d DOS.

%%%%%%%%%%%%%%%%%%%%%%%%%%%%%%%%%%%%%%%%%%%%%%%%%
%%%%%% Figure 2
%%%%%%
%%%%%%%%%%%%%%%%%%%%%%%%%%%%%%%%%%%%%%%%%%%%%%%%%

Provided that the Mn concentration is not too small, the
pseudo-potential on the atomic sites which scatters the valence
electron is mainly due to the Mn atoms themselves (the Al
pseudo-potential being much weaker), and the pseudogap results
mainly from the diffraction of the valence states by the
sub-lattice of the Mn atoms\,\cite{GuyPRB95}. This effect can be
analyzed in term of a long range effective pair potential (up to
$\rm \sim 10\,\AA$) which moreover depends on the magnetic state
of Mn\,\cite{E_paire}. Therefore the magnetic state depends also
on the medium range order of the Mn atoms.

%%%%%%%%%%%%%%%%%%%%%%%%%%%%%%%%%%%%%%%%%%%%%%%%%
%%% Section 4
%%%
\section{LMTO calculations for approximant}

The electronic structure and the magnetic structure of
1/1-approximant and a Taylor phase are performed by using the
self-consistent linear muffin tin orbital method
(LMTO-ASA)\,\cite{art_andersen}.

\begin{figure}[t]
%\vspace{10 cm}
\hfil \psfig{file=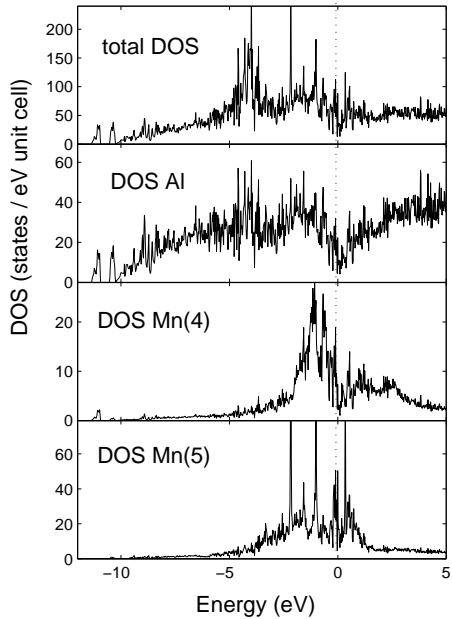,width=6cm} \caption{Electronic
structure of the 1/1-$\rm Al_{65.9}Pd_{12.2}Mn_{14.6}Si_{7.3}$,
calculated by LMTO-ASA without spin polarization: total DOS; sum
of local DOS on Al sites; local DOS on Mn(4); and local DOS on
Mn(5). E$_F=0$.} \label{Fig_1/1-phase}
\end{figure}

We studied the $\rm 1/1-Al_{65.9}Pd_{12.2}Mn_{14.6}Si_{7.3}$
containing 123 atoms in a cubic cell\,\cite{Sugiyama98}. The
lattice parameter is $\rm a=12.281\,\AA$. Two Wyckoff sites (noted
Mn(4) and Mn(5)) are occupied by Mn atoms. The total DOS and the
local DOS on {Al}, Mn(4) and Mn(5), calculated without
spin-polarization, are shown on Fig.\,\ref{Fig_1/1-phase}. The
local DOS on {Al}, that correspond mainly with valence sp states,
exhibits a pseudogap near E$_F$ as expected for an approximant.
The local DOS on the two Mn sites differs markedly. Mn(4) DOS is
split into bonding and anti-bounding states separated by the
pseudogap. In contrast, there is no pseudogap at E$_F$ on Mn(5)
DOS.

%%%%%%%%%%%%%%%%%%%%%%%%%%%%%%%%%%%%%%%%%%%%%%%%%
%%%%%% Figure 3
%%%%%%
%%%%%%%%%%%%%%%%%%%%%%%%%%%%%%%%%%%%%%%%%%%%%%%%%

The spin polarized LMTO calculation converges to a non-magnetic
state ($\rm m=0$) on Mn(4) and on Pd; but there is a localized
magnetic moment on Mn(5), $\rm m=gS\simeq0.7$. Mn(4) is surrounded
by 10 Al atoms, which favors a strong sp-d hybridization and
therefore a non-magnetic states. On the other hand, Mn(5) atom is
surrounded by 5 Mn(5) and 7 Al. The d-d overlapping between Mn(5)
atoms is strong and the sp-d hybridization is less important that
for Mn(4). It follows that Mn(5) in a good candidate to be
magnetic.

\begin{figure}[t]
%\vspace{5 cm}
\hfil \psfig{file=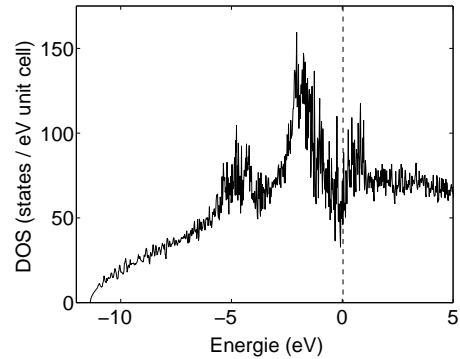,width=6cm} %
\caption{Total DOS of T-$\rm Al_{79.5}Pd_{5.1}Mn_{15.4}$,
calculated by LMTO-ASA. E$_F=0$.} \label{Fig_T-phase}
\end{figure}

The composition domain of the Taylor phase is rather
wide\,\cite{Klein96}. Five Wyckoff sites (noted Mn(1) to Mn(5))
are occupied by Mn only and five Wyckoff sites (noted TM(1) to
TM(5)) by Al/Mn mixture with different Al/Mn ratios. We calculated
the electronic structure for a Mn-low T-phase and we assumed that
TM(1) to TM(5) are occupied by Al atoms only and that TM(6) is
occupied by Pd atoms only. There are 156 atoms in a orthorhombic
unit cell with the composition T-$\rm Al_{79.5}Pd_{5.1}Mn_{15.4}$.
The lattice parameters are assumed to be the same as for a Mn-rich
T-phase: $\rm a=14.717\,\AA$, $\rm b=12.510\,\AA$ and $\rm
c=12.594\,\AA$\,\cite{Klein96}. The self-consistent spin polarized
calculation converges to a spin equal to zero on all Mn and Pd
atoms. The total DOS (Fig.\,\ref{Fig_T-phase}) exhibits a well
pronounced pseudogap near E$_F$ comparable to that found for other
approximants\,\cite{FujiPRL91,Hafner97}. This pseudogap is present
in the local Al DOS, Mn(2) DOS and Mn(5) DOS. Nevertheless it is
not present on local Mn(1), Mn(3) and Mn(4) DOS. The local DOS at
E$_F$ (per atom) on Mn sites are the following: $\rm 1.5
\,states/(eV.atom)$ on Mn(1), $\rm 1.0 $ on Mn(2), $\rm 2.2 $ on
Mn(3), $\rm 2.2 $ on Mn(4) and $\rm 1.1$ on Mn(5). There is a
rather large difference between this DOS; but, all these Mn have a
spin equal to zero. Nevertheless, during the self-consistent
procedure the spin of Mn(3) and Mn(4) converge very slowly to
zero. This suggests that these two atoms are closed to the
magnetic\;/ non-magnetic transition.

Experimentally\;\cite{Francoise_Approx}, the Mn-poor T-phase is
nearly non-magnetic but other T-phases contain magnetic Mn. The
comparison of several T-phases measurements and our calculation
suggests\;\cite{Francoise_Approx} that Mn(1) to Mn(5) atoms are
always non-magnetic and moments are carried by the Mn atoms
occupying the TM sites. This is in agreement with the fact that
the Mn atoms located on the most stable Mn sites are non-magnetic
whereas the Mn atoms located on the less stable sites could be
magnetic as shown in section\,\ref{SecEnergie}.

%%%%%%%%%%%%%%%%%%%%%%%%%%%%%%%%%%%%%%%%%%%%%%%%%
%%%%%% Figure 4
%%%%%%
%%%%%%%%%%%%%%%%%%%%%%%%%%%%%%%%%%%%%%%%%%%%%%%%%

\section{Conclusion}

In Al-Mn-(Si) and Al-Pd-Mn-(Si) quasicrystals and their
approximants many Mn atoms are close to the magnetic /
non-magnetic transition. We show that the magnetic state is
determined by the local DOS on Mn atoms which depends on their
local environment and also on their environment up to $\rm
5-10\AA$. The Mn atoms located on the less stable atomic position
could be magnetic, while the Mn atoms located in the most stable
position are non-magnetic.

%
%  END MAIN TEXT
%

\hp

\begin{footnotesize}
\begin{frenchspacing}

\end{frenchspacing}
\end{footnotesize}

\end{document}